# Developing a Compiler for EROP - A Language for the Specification of Smart Contracts, An Experience Report

Adrian Delchev, Ioannis Sfyrakis[1], and Ellis Solaiman[1]

School of Computing, Newcastle University, UK
ioannis.sfyrakis@newcastle.ac.uk, ellis.solaiman@newcastle.ac.uk

Abstract. A smart contract is a translation of a standard paper-based contract that can be enforced and executed by a contract management system. At a high level of abstraction, a contract is only a document that describes how the signing parties are to behave in different scenarios; nevertheless, the translation of a typical paper-based contract to its electronic counterpart has proved to be both time-consuming and difficult. The requirement for a language capable of capturing the core of a contract in simple phrases and definitions has been a focus of study for many years. EROP (Events, Rights, Obligations, Prohibitions) is a contract specification language that breaks a contract down into sets of events, rights, obligations, and prohibitions.

## 1 Introduction

Both businesses and academia require novel methods of automating the process of smart contract compliance monitoring [1] [2]. A well-designed contract management system may significantly minimise the human component of contract compliance monitoring and verify that the business operations of partners adhere to the terms of the contract being enforced. A management system is required to monitor the interactions between parties to a contract to ensure that their company operations and activities are carried out appropriately and in accordance with the contract's terms. The CCC (Contract Compliance Checker),an impartial, third-party monitoring service developed at Newcastle University [3] [4] is an example of such a service. The system itself was designed with a conceptual language for writing smart contracts called EROP. EROP is a contract definition language that uses JBoss Rules, sometimes known as Drools, for rule administration. EROP stands for events, rights, obligations, and prohibitions [5].

The implementation of the CCC relies on the EROP ontology - a set of concepts and relationships used for modeling the execution of business operations between partners, and for reasoning about the compliance of their actions. The EROP ontology is implemented in JAVA as an extension to the Drools engine, which allows for a better, more direct mapping of the EROP specification language to the concrete implementation. Having the option to express a contract in the EROP language allows for a broader user base since only limited technical knowledge will be required to convey a contract in EROP as opposed to writing it in the extended Drools directly for monitoring. As it currently is, a translation from EROP to the extended version of Drools (also known as Augmented Drools or AD) is a manual process, which requires the contract specification in EROP to be mapped to its AD equivalent. The main contribution of this paper is to design and implement a translation engine that would automate the process of EROP to AD translation and provides a valid and correct output for an input that complies with the specifications of the EROP language. We measure the effectiveness of the solution by comparing previously confirmed and verified manual translations from EROP to AD. Our solution has been designed with a modular structure encapsulating required functionality and allowing for easier maintenance and enhancement in the future if needed. A translation engine for EROP to AD mapping eliminates the need for manual translation, potentially significantly reducing translation and mapping related errors, and better utilization of time spent expressing a paper contract as its electronic smart contract equivalent.

## 2   Smart contracts: background

### 2.1   Contract specification languages

The presence of contracts in our society is ubiquitous from every day simple oral agreements that we take for granted to formally specified and notarized documents that have strong and profound effects on our lives. Their wide use today is undeniable and its roots can be traced back to ancient societies [6]. The advancement of electronic commerce has increased the sheer number of contracts an organization can take part in, resulting in difficulties in keeping up with the requirements of an electronic market. Creating a contract is a task that requires significant resources and efforts from hiring adept personnel to formally specify and verify the contract parameters to negotiation between parties and mediation with the business. Electronic smart contracting aims to automate the process of contract establishment and execution while reducing development costs. To this end, a contract has to captured in a language that has the expressive power to specify all the contract content while eliminating any ambiguities [7] [8] [9] [4] [10]. In order for a contract to be eligible to be monitored and enforced by a contract compliance system of any kind it has to be formally specified in a language that has the ability to capture the requirements of a contract including legal requirements, clauses and internal policies as well as the acting parties and their actions. Given that the desired outcome of a contract monitoring and compliance service is automation and execution of contracts with minimal human interference, the language that captures the contract has to be precise and free of ambiguities so that the need of manual conflict resolution does not arise. Smart contract research has gained significant momentum since the rise of Blockchain technology with the development of Bitcoin, Ethereum, Hyperledger, and other blockchain technologies. Indeed smart contracts have been found useful in a range of application areas including education [11] [12] [13] [14], Internet of Things [15] [16], and Service Level Agreements (SLAs) [17] [18] [19] [20] [21]. The idea of smart contracts, however, has been around much earlier than the development of blockchain, and can be implemented on a host of centralised, distributed, and hybrid architectures [22] [23] [24]. Research into smart contracts and blockchain covers a range of areas including performance [25] and simulation [26] [27].

The proposed contract representation language called DocLog [28] introduces the concept of legal advice systems capable of providing, while not as sophisticated as an actual experienced law professional, legal advice about exchanged messages. DocLog uses a tri-layer structure that combines data, text and semantic oriented approaches for exchanging contract terms. The data layer aims to provide the contract data in such a way that it can easily and efficiently be processed by a transaction processing system. The text approach is represented by a Natural language layer that strives to present the contract terms in a way that is easily comprehensible by human users. It uses XML to structure the content of a contract which allows the support for individual clauses, sub clauses, sections, sub sections and allows the use of version and approval management systems. While DocLog does provide a relatively human readable way of capturing contract specifications, it does not provide any means of monitoring the captured the contract or allow any manipulations on the captured properties of the contract. Furthermore several important aspects of contract specifications cannot be captured using the DocLog language. Temporal aspects such as Buyer has to submit payment no later than 5 days after making an order as well as exceptional circumstances within contracts that specify what is to be done when an aberration from the norm of contract occurs are not presented in the original version of DocLog. Nevertheless the underlying architecture of the language provides insights on communication between different layers of the language and mapping between layers using the EDI Translator [29].

The presented administrative architecture [30] supports the management of e-contracts and defines the ontology used in a business partnership for the underpinned contract. This is covered in four steps: 1) Consistency based off-line verification and achievability of contract aims given the possible reachable system states. 2) Definition and compilation of the application specific processes that facilitate the execution of the contracts such as enactment, monitoring, updating, termination, renewal etc. 3) Definition of the roles of the interaction agents. 4) Identifying different components and services used by acting agents necessary for them to play their respective roles in the partnership. The main focus of the framework as discussed in [31] is centered around corrective monitoring , where violation of contractual norms are detected and then tried to be fixed using corrective measures as opposed to predictive monitoring where such violations are predicted by using the agents behavior and actions are taken in order to completely avoid undesired behavior. The actual contracts are captured in an XML based language with a multi-layered architecture. The language is capable of expressing different contract structures such as clauses, parties, groups and actions using deontic notions such as obligations, permissions and prohibitions and even though it uses a relatively high level declarative style of writing its implementability is unclear. It is possible to represent concepts, intuitively acceptable to humans but unfortunately it is not possible to unambiguously translate them in a from that would allow it to be processed by machines. Unlike DocLog, the language offers a degree of exception handling; however it appears to be limited and is not the main focus of the project and neither is usability.

There is a myriad of other options when it comes to contract specification but unfortunately most of the available languages either do not focus on usability and practicality or dont have a contract monitoring system in which they can be integrated with ease. BPEL [32] uses event calculus to represent actions and their corresponding effects and is not targeted towards business contracts but for specifying web service interactions. It offers a less declarative approach than the other presented languages. Exception handing is present but recovery from exceptional cases is limited the biggest issue seems to be usability. The abstract notation may be too daunting for non-technical personnel to use for commercial purposes. Heimdahl [33] is a platform for monitoring obligation policies and it uses xSPL [34] for the specification of those policies. xSPL syntax is declarative but some of the language constructs are not intuitive and might be difficult to understand for people without a technical background. Exception handling also does not appear to be possible directly.

2.2  EROP, Augmented Drools, and the CCC

As introduced in [5] EROP is a contract specification language that focuses on execution and business resolution of a business partnership. The language relies heavily on events, rights, obligations, and prohibitions, and it captures the information of a conventional paper-based contract into sets of the aforementioned constructs. One of the most significant additions that distinguish EROP from other contract specification languages is the extended capabilities that allow for reasoning and providing resolution of unforeseen circumstances that arise from business and technical failures. One of the strengths of the language stem from the fact that at its level of operation the low level details are abstracted away, allowing contract writers to concentrate on expressing the business operations of a contract. Another selling point of the language is its ease of use even for non-technical personnel and the already existing implementation of its ontology that allows for contractual compliance monitoring.

A contract written in EROP consists of two sections a declaration section where all the acting role players, business operations and composite obligations used in the rules are defined, and a rule section that captures operations and manipulations of the entities specified in the contract as well as any actions and exceptional behavior.

The formal grammar of the language as well as more details on the EROP syntax will be provided in a later section. The EROP ontology as specified in [5] is a set of concepts and their relationships within the domain of B2B interaction that we employ to model the evolution of interactions between business partners, for the purpose of reasoning about the compliance of their actions with their stated objectives in their agreements. The ideas of the EROP ontology have been implemented as set of Java classes that capture the properties and available operations on those properties; the implementation extends the rule language offered by the Drools rule engine, adding various different construct to reason about and manipulate the operations of business partners within a contract. The implementation of the ontology, also known as Augmented Drools is less abstract and readable than EROP and closer to Java in style. It also needs additional code for convenience and housekeeping purposes that are needed for the implementation of the ontology to work but bot necessary for human reader, initializing a contract in EROP. The EROP language maps completely to Augmented Drools, which makes it possible for direct language to language mapping. In this sense, the problem of creating a precise and formal grammar of the EROP language becomes one of translating EROP to Augmented Drools, which is the main topic of this work and will be explored in detail in the upcoming sections.

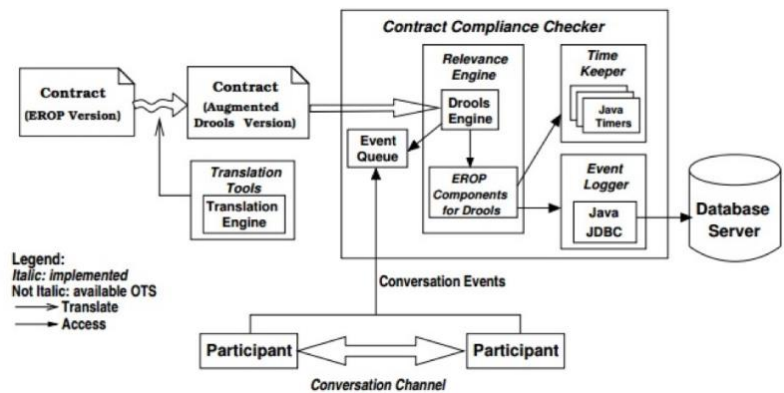

Fig.1. Representation of the CCC architecture [5]

The Contract Compliance Checker - for short the CCC is the contract compliance monitoring service as introduced in [3] [4]. It is a neutral entity conceptually standing between the interacting parties and its purpose is to monitor the exchange of events between participating entities and infer whether or not the business operations these events relate to are compliant or non- compliant to a specified contract. The architecture of the CCC is illustrated in Fig. 1 but any further discussion on the inner workings of the CCC will be limited as its not the topic of the work presented here. For the purposes of this project it is important to note that all smart contracts expressed in EROP or Augmented Drools can be uploaded to the CCC for contract compliance monitoring.

2.3     EROP to Augmented Drools translation

As mentioned in the previous section EROP maps directly to Augmented Drools given that the former is derived from the latter. In this case a translation between EROP to Augmented Drools essentially comes down to automating the mapping between the two languages. The techniques common for most types of translations will be discussed and reviewed in detail in the next section, here we give some of the more specific techniques available.

As discussed in [30], the translation of rules expressed in language based on natural or close to natural format to a rule standard can be accomplished by two-fold mapping where the extracted natural language is mapped to Java beans, which serve as intermediary for the translation between the source language and JBoss Drools production rules. The translation technique uses a straight forward verb and noun concepts grammar that captures the parameters of the source language and is then mapped into rules. Although the paper describes natural or close to natural language mapping, which requires more effort in comparison to EROP to AD mapping because of the lack of rule structure of the source language, it is interesting to note that an intermediary layer in the face of Java Beans is used in order to capture the required information to build an operational rule. The Authors of [35] discuss the approach used by Object Oriented rules systems such as Drools and ILOG Jrules. Such systems are built on top of Java vocabularies in the case of Drools, Java beans are used as facts to represent the domain of the rules and their vocabulary in user applications. Different vocabularies are used by rules through the import declaration, specified inside of the rule file. The paper describes the approach used to translate from Drools using the low level structure of the language such as beans to translate rules to R2ML, which is an XML based Rule Markup Language. Even though the direction of translation described in the paper goes in the opposite direction of the one desired from the EROP to AD translator, it does highlight the importance and benefits of using an XML structure or language for translation to Drools.

2.4     Summary

The conducted research demonstrates that among the presented contract specification languages, none has the expressive power or the ability to deal with exceptional circumstances arising from business or technical failures as EROP does. All of the presented languages require an extensive technical background and may be daunting for a non-technical person to use. In addition to that some of them are merely a notation for expressing contracts, with no definitive means of monitoring the expressed contract for compliance. EROP along with the CCC represent a complete solution for capturing contract requirements and then monitoring and enforcing them. A translator from EROP to AD has to be based on direct mapping and may make use of intermediary representations to hold its data. As presented in the reviewed approaches, JavaBeanslike structures and XML are capable of accomplishing the desired goal and have been used by other similar projects. The next section covers some of the fundamentals in translation and how they are applied in the development of the EROP to AD translator.

## 3 Methodology

The development of an EROP to Augmented Drools translator is a task that requires extensive analysis of current techniques and translation technologies. Our initial aim was to approach the development process incrementally, dividing the overall project into several specific sub projects that would be united in the end to produce the final solution. During the initial stages of the development process the focus was on analysis and comprehension of the state of the art of the translation scene. The analysis revealed the parts of common translation techniques that would be essential for developing a translation and also gave an insight into the architectural design of the solution. The use of external, third parties' libraries was taken into consideration and the necessity of such has been reviewed in greater detail later in this section. The programming language used for the development of the solution is JAVA. As its implementation dependencies have been reduced to a minimum and it provides the required functionality to achieve the task at hand. The target of the translation the implementation of the EROP ontology is also in JAVA, which makes the choice to use the language as a logical one as it will promote consistency throughout the contract compliance monitoring solution [5] developed at Newcastle University.

### 3.1 Compiler Analysis

Generally, the purpose of programming languages is two-fold they serve as a notation for describing computations to both machines and people. Other than formally expressing a programmer's intention, they exist for the purpose of bridging the gap between different layers of abstractions the higher layers that are easier to comprehend and safer to use, and lower levels that are often more efficient and flexible. A language that is more human oriented needs to be translated to a language that a computer understands. Such translation software is known as a compiler. A typical compiler breaks the mapping of a source language to the target language in several stages. Most commonly the first of those is the analysis stage it breaks a source program into pieces and verifies the grammatical structure of the source language on them. The resulting pieces are then used for the creation of an intermediary representation of the source language. The analysis stages duty is to detect syntactical and semantical compliance or inconsistency. The intermediate representation built by the analysis stage is called a symbol table and after its creation it is then passed to the next stage of the process. The second stage is the synthesis, where a translation to the target language is created using the intermediary representation of the source language. In a sense the analysis part is the front end, and the synthesis is the back end of a compiler. When reviewed in more detail, the processes of a translation are executed in a sequence of steps (fig2)

The initial step of the translation process is carried out by the Lexical analyzer. Its purpose is to read the characters making up a source program or a file and use them to create meaningful sequences called lexemes, producing tokens containing the parsed information, which is then used in the syntax analysis stage. To put things into context an input in the form of the EROP language such as:

POAcceptance     in     buyer .rights

Would be broken down by a lexical analyzer into the following lexemes: POAcceptance is a lexeme that would be mapped to the token (id,1), where the 1 is pointing to the position of PAAcceptance in the generated symbol table containing information such as value and type. in and. Would be mapped to the tokens (in) and (.) because they are both operations of the language Similarly to POAcceptance, buyer and rights will be mapped to (id,2),(id,3) And the resulting token mapping would be:

( id , 1 )    ( in )    ( id , 2 )    ( . )    ( id , 3 )

The next step of a translation process within a compiled is the syntax analysis, also known as parsing. It uses the tokens created by the lexical analyzer to create a tree like intermediate representation of grammatical structure of the source language. Most commonly [19] the intermediary representation is known as a syntax tree in which each parent node represents an operation and the children nodes of the parent represent the arguments needed to complete the operation. Using the above breakdown of input characters into tokens, the resulting syntax tree for the given input would look like:

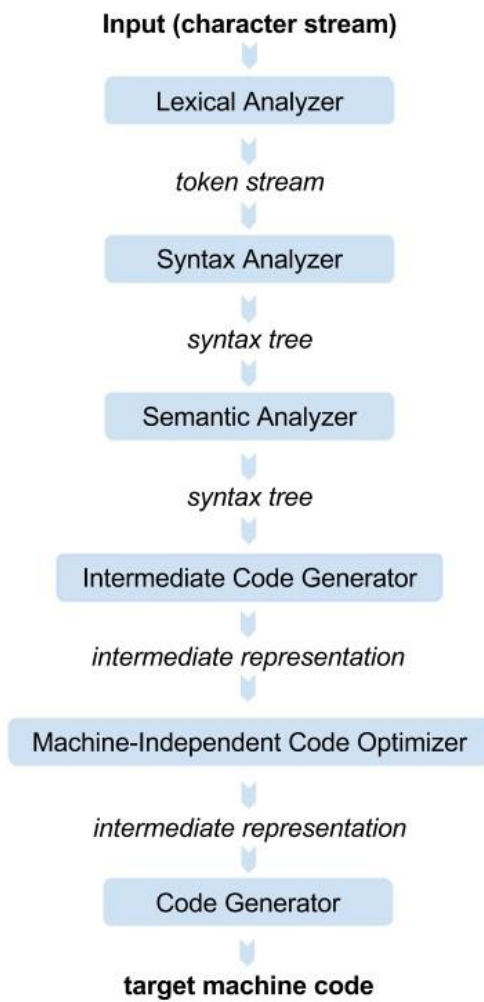

Fig.2. Common phases of compiler translation

The node labelled as dot indicates that the operation must be completed using the children nodes and that the produced result should serve as the right hand side of the operation labelled in. The next step of the process is the semantic analyzer. It uses the information stored in the symbol table as well as the generated by the syntax analyzer tree in order to check that the source program complies semantically with the definition of the language. One of the most important parts of the semantic analysis is type checking in other words where the operands are from the appropriate type for the specified operator. For instance, in most programming languages an array is indexed using an integer value, if the compiler detects anything that does not match the expected type it is supposed to notify the user for the inconsistency, in the case of EROP an example would be the definition of a business operation. As stated in [5] a business operation is defined by a generic string that starts with an upper letter. In that sense a string with a lower letter would be the wrong type when trying to define a business operation and therefore the compiler would have to return an error. The step following the semantic analysis is the intermediate code generation. The intermediary translation can be more than one or it can be expressed in a variety of different forms. The most important properties of the intermediate representation are that it should be easy to produce, and it should be easy to translate into the target language. The final step of the compilers process is the code generation, which uses as an input the intermediate representation of the source program and maps that to the target language, where different approaches are used depending on the differences or similarities between the source and target languages. The detailed view of a compiler inner workings have been invaluable to the development and design process of the final solution. It has contributed towards the understanding of how typical language translators work and what the most common aspects in them are. Given the semantic similarities between EROP and Augmented Drools (the former is derived from the latter) a translation process between the two languages doesn't need to include all of the steps undertaken by a typical compiler, namely the code optimization step is redundant. Furthermore, in order to speed up the development process and avoid needless low level errors oversights, specialized tools that have efficiently implemented some of the outlined principles can be used.

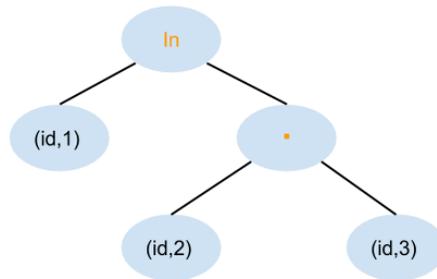

Fig.3. A Syntax Tree

## 3.2 Parser Generators

Parser generators are a specific class of software development tools that are able to generate the framework needed for a program to implement a parser from a set of rules called grammar. A few different methods exist for parsing a given stream of character input, but the two most important ones are top-down and bottom up analysis algorithms as they apply to the widest range of input grammars and context-free grammars and are appropriate to use in a parser generator. Tools such as parser generators have been used in the creation on compilers and translators can be traced back to the early days of computing with examples dating back to 1965 [36]. The two major advantages of using parser generators are firstly development time once proficient in writing grammars using a generated lexer and parser expedites the development process immensely. The second advantage of parser generators is correctness by construction, meaning that the generated parser accepts exactly the language specified in the grammar used to create it [36]. Today a wide range of parser generators exist, with all of them employing similar input parsing techniques and differing from one another in terms of style of grammar specification, algorithms used to parse the input, language of the generated parser files and so on. As mentioned in the development of EROP, a parser generator called ANTLR was considered when thinking about the translation between EROP and Augmented Drools [5].

ANTLR is a parser generator that has a wide range of uses including reading, processing, executing or translating structured text or binary files. ANTLR [37] supports actions and attributes flexibility, meaning that different actions can be defined in separate files from the grammar and essentially decoupling it from the target language, enabling easier targeting of multiple languages. ANTLR can also be used to generate tree parsers and processors of abstract syntax trees. It uses EBNF as a format of its grammar input and has support for popular IDEs. An additional benefit is that it generates a lexer as well as a parser and the resulting generated files are in JAVA format, which makes it consistent with AD and the language used for the development of the EROP to AD translator. ANTLR is also widely popular and is used by Twitter for query parsing, processing over 2 billion queries a day as well as in projects such as Groovy, Hibernate, IntelliJ IDEA and many more.

## 3.3 Solution Architecture

After reviewing the most common translation techniques and approaches, the architectural design of the EROP to Augmented Drools translator started to emerge. The use of parser generator would enable a more rapid development and allow to some extent to reuse the formal grammar of EROP as specified in [5]. The use of ANTLR would also mean that the first three steps of typical compiler architecture can be accounted for and there is no need to create spate entities for the desired functionality. As noted earlier, machine independent code optimizer would not be practical because of the similarities of the target and source languages, which essentially means that the intermediary code generator can interact directly with the code generator where the mapping is made, and the final result is produced. As discussed in a previous section an intermediary layer can be accomplished by an implementation like JavaBeans. The disadvantages of using JavaBeans directly are that it supplies nullary constructors for all its subclasses, which means that they are at risk of being instantiated in an invalid state. The problem stems from the fact that a compiler cannot detect such instantiation which can lead to troublesome debugging and tracing, especially when using a generated parser. Nevertheless, in order to accommodate the intermediary code generation layer of the EROP to Augmented Drools translator, similar in concept Collection of Java classes can be implemented that captures the essence of the building blocks of the EROP language.

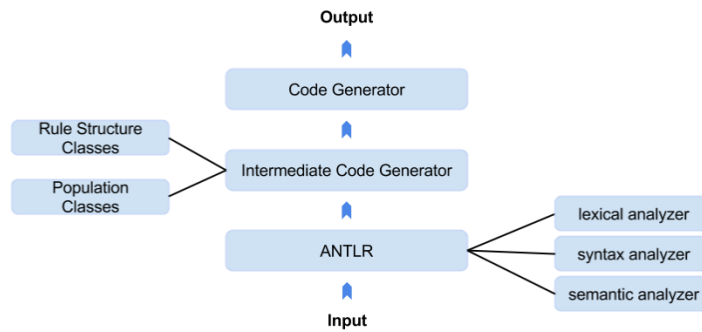

Fig.4. The Conceptual architectural design of an EROP to AD translator

The collection of those classes, well as any additional classes required to accommodate communication between different layers of the architecture, will be discussed in the implementation section. The initial conceptual design of the translator is depicted in Figure4.

## 4 Implementation

### 4.1 ANTLR Grammar

A grammar file in ANTLR is simply a file that specifies the syntax and different constructs of the language and how they connect with one another. On a high level of abstraction, the grammar consists of lexer and parser rules that once specified are then embedded in the generated by ANTLR lexical and syntactic analyzers. The lexer rules begin with an upper-case letter, as opposed to the parser rules and are used to tokenize the input. Lexer rules are essentially the fundamental, building blocks of a language. Parser rules on the other hand are more complex rules that can contain rules themselves as well as tokens characterized as fundamental to the Language. As specified in [3], a contract expressed in the EROP language consists of two parts a declaration section, where all the role players and business operations are defined and a rule section, where the different rules used for compliance monitoring are captured. From that we can infer that the root structure of the language is a contract file and everything else is contained within that file. That can be represented in ANTLR as the entry point of any received input, and it would have any number of children depending on what a contract file can contain and what different constructs in the contract file can contain themselves. To put things into context, Fig 5 gives a partial visual representation of the structure of a contract file defined in the EROP language.

As depicted in the partial representation, a declaration section can have one or more declarations and each declaration is a business operation, role player or composite obligation declaration. The role player declarations as well as the identifier are specified at the lowest level to give a feeling of what rules, sitting at the bottom of the rule hierarchy would look like. A role player declaration simply consists of the keyword Roleplayer followed by white space and one or more identifiers and ending with a semicolon. The one or more quantifier makes it possible to declare multiple role players in a single line. An identifier is simply a string starting with lower case and the ability to contain uppercase letters as well as digits.

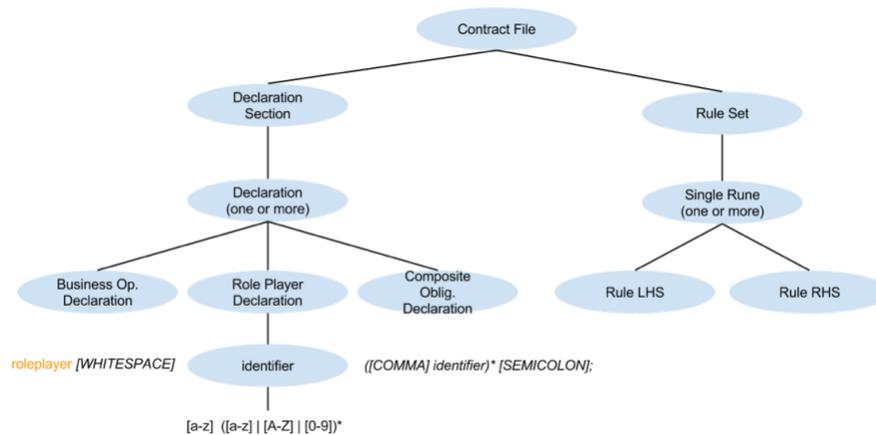

Fig.5. Partial representation of EROP grammar.

A grammar is an important part of ANTLR, but by itself it does not provide much functionality because the associated parser is only able to tell us whether an input conforms to the language specification given. In order to build translation or any type of applications for that matter, there is a need for the parser to trigger some sort of action, whenever it encounters input sequences, phrases or tokens of interest. Fortunately, ANTLR provides two mechanisms that allow invocation of actions it automatically generates parse tree listeners and visitors to enable building language applications. A listener is an object that can respond whenever it detects rule entry and exit events triggered by a parse tree walker as it discovers and finishes nodes which means that ANTLR automatically generates the interfaces for any entry or exit events. The most profound difference between listeners and visitors is that listener methods do not have the obligation to explicitly call methods to walk their children that gives flexibility in a scenario where only specific parts of the input language should trigger events. The alternative is visitors they must explicitly activate visits to their child nodes in order to keep the traversal of the tree going. In the case of an EROP to Augmented Drools translator the alternative method makes much more sense as parsing of all the input information is needed. The alternative is visitors they must explicitly activate visits to their child nodes in order to keep the traversal of the tree going. In the case of an EROP to Augmented Drools translator the alternative method makes much more sense as parsing of all the input information is needed.

The provided functionality allows for triggering specific events when a rule from the grammar is entered. There is still the need for implementing specific actions when such events occur. Given that re-usability is an important part of the software development process it would make sense to reuse common concepts instead of creating duplication. Identifier is an example of commonly repeating grammar structure that can be reused. It is used to describe business operations, composite obligations as well as role players, not to mention that it can occur not only in the declaration section but also in the rule set when role players and business operations are referenced, or their ROP sets manipulated. In order to allow re-usability while keeping the ability to distinguish for which part of the grammar common grammatical structures refer to Ive implemented two additional JAVA classes that serve as a buffer for the ANTLR parser and population classes that create the intermediary representation of the EROP language. Those classes are Variables Flagger and Variables memory. Their purpose is to respectively activate various flags whenever the ANTLR tree walker enters different rules and then use those flags in order to make decisions on where the contents of the parsed file should be stored. The separation of communication between the ANTLR parser and the intermediary representation of EROP follows good software development guidelines and practices as it encapsulates and abstracts away the logic needed to make the decisions about where the parsed information goes. It also helps with testing and contributes to the modular approach design, which is one of the development aims of the project.

4.2    The Rule Structure Classes

As discussed in the analysis section, the intermediate code generator of the translator can be accomplished by custom Java classes inspired by various techniques. The resulting Java classes would have to capture the structure of the corresponding language constructs and any information they hold that are needed for linking the intermediary representation to the final target language. The resulting classes and their functions are as follows:

- EventMatchCondition: represents the conditions an event match has to satisfy in order for the event to be triggered. It follows the structure field operator value and as specified in the full language grammar (included in the appendix) the field value can be any one from botype/outcome/originator/responder as specified in [3].
- Constraint: the constraint class is a generalized collection of any constraints that can be specified on a rule. It can hold any of the following: RopConstraint capturing the presence of absence of particular business operations or composite obligations in a role player ROP sets; HistoricalConstraint used to condition the triggering of rules depending on the presence or absence of certain events or the times a specified event occurred; TimeDirect and TimePartialComparisons constraints used to enforce additional checks on the timestamp of a given event; TimeDirectComparison is used to check of a timestamp the same as, before or after a specified point in time; TimePartialComparison is used to check if an event timestamp is within a given range of hours, minutes, years, days or months; OutcomeConstraint used to specify a constraint on the outcome of an event such as Success, Fail, BizFail etc; RhsAction represents the right hand side of a rule, anything between the then and end part of a rule. It can contain conditional statement, outcome or pass actions as well as any manipulation on a role players ROP set or the outcomes of a business operation.
- IfStatement: a conditional structure that is used to capture additional constraints in the RHS of a rule. It comes with its own left- and right-hand side and even though it doesnt alter the EROP languages structure it allows for a more natural and productive style of writing.
- AddOrRemAction: used to gather information about any manipulation of a role player ROP sets such as adding or removing Rights/Obligations/Prohibitions.
- Rule: the root class in the rule class structure architecture that contains all the information required to represent and recreate a rule.

Figure 6 shows the architectural hierarchy of the classes in the Rule Structure and how they interact with one another.

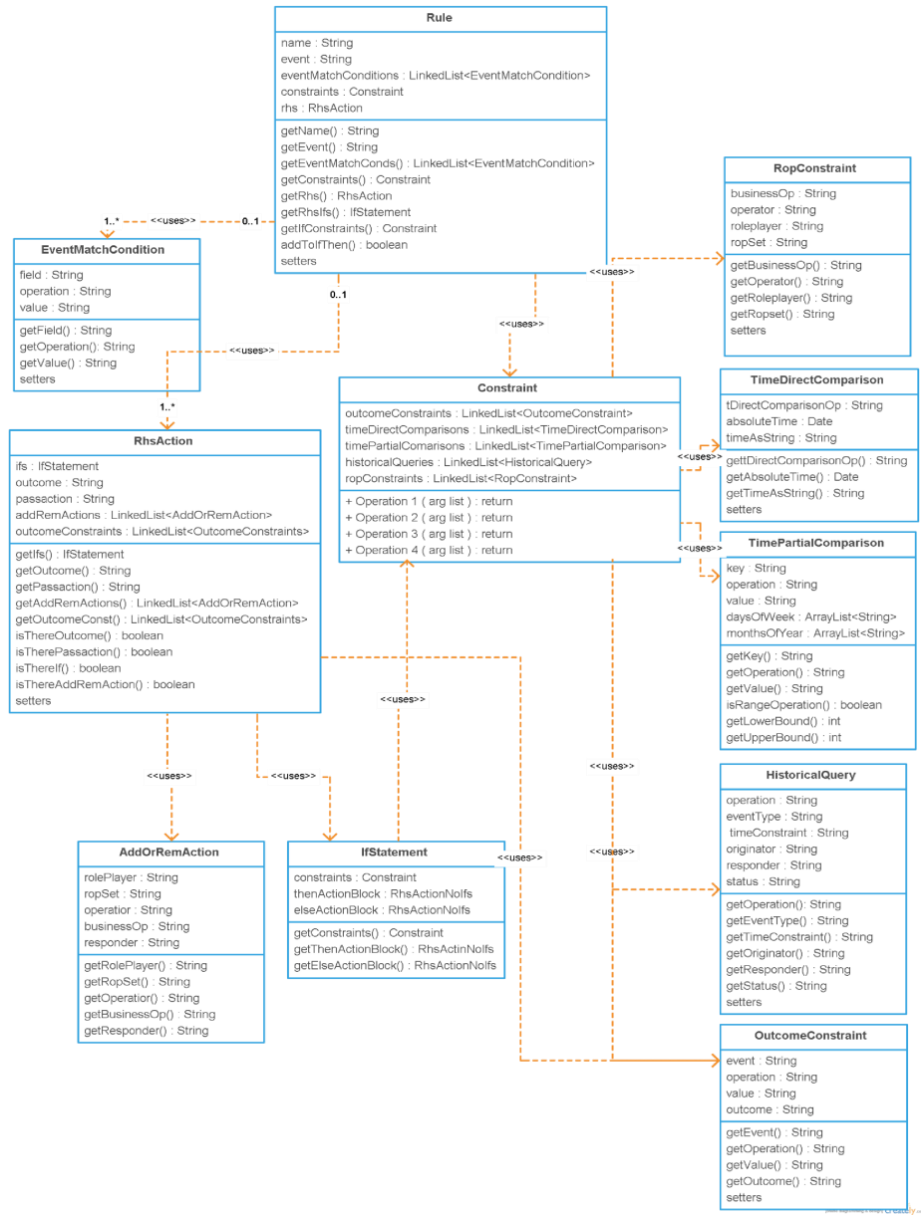

Fig.6. UML Diagram

```
rule "Rule1"                                    rule "Rule1IfThen"
    when e matches (eventMatchConds)                when $e: Event (eventMatchConds)
    then                                                eval (booleanConditions)
        if (booleanConditions)                      then
        then                                            actionBlock1
            actionBlock1                        end
        else
            actionBlock2                        rule "Rule1IfElse"
end                                                 when $e: Event (eventMatchConds)
                                                        eval ( ! booleanConditions)
                                                    then
                                                        actionBlock2
                                                end
```

Fig.7.

### 4.3  Lookup and Mapping

With a suitable ANTLR grammar and an intermediate Structure to represent EROP, the only thing left to make a working translator is a Mapper. The Mapper Functionality is executed by the Translator Java Class. Its purpose intuitively is to translate different parts of the rules from the intermediary format to the final target language Augmented Drools. Given that the intermediary format is quite close to the target Language a direct mapping, with some decision making is a suitable option for the translation. The Translator also includes decision making logic that allow it to determine whether a single rule in the source language needs to be broken down into multiple ones in Augmented Drools. This is most commonly due to conditional statements in the rule structure of the source language and is since conditional statements do not exist in Augmented Drools. A quick example to illustrate the mapping process is given in figure 7.

In general, a rule in EROP containing an if else statement maps to two rules in Augmented Drools one with the if condition added to the when condition set in AD and the then action added to the right-hand side of the AD rule. The second rule contains the negated if condition to the when condition set, and the else action added to the right-hand side of the AD rule. It's worth noting that the second rule only needs to be generated if there is an else action block, in the case where there is only an if action block, the second rule does not need to be translated. The final piece of the translator is the Lookup, implemented as a Java Class using an XML file containing the corresponding mappings of methods of the EROP Ontology classes to the keywords in EROP. The design is inspired by the various XML languages used as an intermediary layer as discussed in the previous sections and is also quite practical because it encapsulates all the mapping to a single file. In the case when the name of a method is changed in Augmented Drools, the only change needed to be made to the Translator is in the lookup file.

## 5 Evaluation

In order to critically examine and determine the correctness of the developed tool as well as any addition, enhancements and/or omissions that need to be made to both the EROP language or the Augmented Drools implementation, we present a case study in which the concepts of a contract represented in the latest version of Augmented Drools are extracted and serve as key points that are then used to express the same contract in EROP. Once the contract is in EROP it can then be ran through the translator and the produced translation can be evaluated against the original contract, expressed in Augmented Drools. The two main aspects on which the translation will be judged are ability to express concepts and correctness of the produced translation with regards to the original of the contract. The contract used for the case study is between two parties, which will be referred to by simply Buyer and Seller. The original contract in its entirety, expressed in Augmented Drools can be found in the appendix section. The clauses of the contract extracted from the original are as follows: C1: The buyer has the right to submit a buy request, having sent a buy request, a buyers right so submit any further ones is revoked until the current one is resolved. At the same time, the seller gains an obligation to either accept or reject the received buy request. C2: In the event of one or more business failures during the buy request, the first business failure should be noted, and any further business failures should reset the Rop sets of the role players. C3: Having received a rejection of a buy request, the pending obligation is satisfied, and the buyer can have its right to send additional buy requests restored. C4: In the event of one or more business failures during the rejection of the buy request, the first business failure should be noted, and any further business failures should reset the Rop sets of the role players. C5: After receiving an acceptance of a buy request from the seller, the pending obligation has been satisfied and the buyer receives a new obligation to pay the seller as well as the right to cancel the order. C6: In the event of one or more business failures during the acceptance of the buy request, the first business failure should be noted, and any further business failures should reset the Rop sets of the role players. C7: After a payment has been received, the buyer has satisfied its obligation; he loses the obligation to pay as well as the right for a cancellation and regains his right to submit further buy requests. C8: In the event of one or more business failures during payment of the buy request, the first business failure should be noted, and any further business failures should reset the Rop sets of the role players. C9: After the buyer sends a cancellation, he loses the obligation to pay and the right to submit further cancellations. C10: In the event of one or more business failures during cancellation of the buy request,

```
roleplayer buyer, seller;
businessoperation BuyRequest, Payment, BuyConfirm, BuyReject, Cancelation;
compoblig ReactToBuyRequest(BuyConfirm, BuyReject)
```

Fig.8.

```
rule "BuyRequestReceived"
      when e matches (botype == BUYREQ,originator == buyer,responder ==
store,outcome == success)
            BuyRequest in buyer.rights
      then
            buyer.rights -= BuyRequest(seller)
            seller.obligs += ReactToBuyRequest(buyer,"01-01-2016 12:00:00")
end
```

Fig.9.

```
rule "BuyRequestBnessFailure"
      when e matches (botype == BUYREQ,originator == buyer,responder ==
store,outcome == tecFail)
            BuyRequest in buyer.rights
      then
            if (BuyRequest.BizFail == false)
                  then BuyRequest.BizFail == true
                  else reset buyer
                        reset seller
            endif
end
```

Fig.10.

the first business failure should be noted, and any further business failures should reset the Rop sets of the role players. Two role players are defined in the contract along with the following business operations: BuyRequest, Payment, BuyConfirm, BuyReject, and Cancelation. The clauses of the contract as specified above, define how the ROP sets of the roleplayers change during the interaction. A file in EROP starts with the definition of the role players, business operations and composite obligations used in the contract.

The second part of the contract contains definition of the rules of the specified roleplayers and interactions between them. The rule for a received Buy Request can be derived from C1 of the contract. It occurs when a successful buy request is received.

The second rule can be derived by C2 and as specified it requires specific actions to be executed depending on a certain condition. The rule can be modelled using a conditional structure in EROP.

The third rule of the contract is derived from C3 and is triggered whenever the seller rejects a buy request from the buyer.

```
rule "BuyRequestRejected"
      when e matches (botype == BUYREJ,originator == store,responder ==
buyer,outcome == success)
            ReactToBuyRequest in seller.obligs
      then
            seller.obligs -= ReactToBuyRequest(buyer)
end
```

Fig.11.

```
rule "BuyRequestRejectedFailures"
      when e matches (botype == BUYREJ,originator == store,responder ==
buyer,outcome == tecFail)
            ReactToBuyRequest in seller.obligs
      then
            if (BuyConfirm.BizFail == false)
                  then BuyConfirm.BizFail == true
                  else reset buyer
                        reset seller
            endif
end
```

Fig.12.

```
rule "BuyRequestConfirmation"
      when e matches (botype == BUYCONF,originator == seller,responder ==
buyer,outcome == success)
            ReactToBuyRequest in buyer.obligs
      then
            seller.obligs -= ReactToBuyRequest(buyer)
            buyer.obligs -= Payment(seller)
            buyer.rights -= Cancellation(seller)
end
```

Fig.13.

The fourth rule of the contract is directly derived from C4 and is very similar in definition to BuyRequestBnessFailure. It occurs when a business failure occurs during a rejection of a buy request.

The fifth rule defines what happens when a successful confirmation of the buy request is received, it is derived from C5.

The sixth rule, similarly to the second and fourth rules describes what happens in the event of failures during the confirmation.

The seventh rule, derived from C7, captures what occurs in the event of a successful payment.

The eight rule describes what happens in the event of exceptional circumstances during receiving a payment, it is derived from C8.

The ninth rule captures what happens whenever a cancellation is received.

```
rule "BuyRequestConfirmationFailuress"
    when e matches (botype == BUYCONF,originator == seller,responder ==
buyer,outcome == tecFail)
         ReactToBuyRequest in seller.obligs
    then
         if (BuyConfirm.BizFail == false)
              then BuyConfirm.BizFail == true
              else reset buyer
                   reset seller
         endif
end
```

Fig.14.

```
rule "PaymentReceived"
    when e matches (botype == BUYPAY,originator == buyer,responder ==
store,outcome == success)
         Payment in buyer.obligs
    then
         buyer.obligs -= Payment(seller)
         buyer.rights -= Cancellation(seller)
end
```

Fig.15.

```
rule "PaymentReceivedBFailures"
    when e matches (botype == BUYPAY,originator == buyer,responder ==
store,outcome == tecFail)
         Payment in buyer.obligs
    then
         if (Payment.BizFail == false)
              then Payment.BizFail == true
              else reset buyer
                   reset seller
         endif
end
```

Fig.16.

The last rule of the contract expresses what is to happen whenever exceptional circumstances occur during the a buy cancellation.

5.1   Outcome and Role player Constraints

We will start be discussing the ability of EROP to express concepts present in the latest version of Augmented Drools as seen in the original of the contract presented in the appendix section. Clauses 2, 4, 6, 8 and 10 of the contract require the ability to check if a certain business action has been set as a business failure, this can be characterized as

```
rule "BuyCancellation"
      when e matches (botype == BUYCANC,originator == buyer,responder ==
store,outcome == success)
            Cancelation in buyer.rights
      then
            buyer.rights -= Cancellation(seller)
            buyer.obligs -= Payment(seller)
end
```

Fig.17.

```
rule "CancellationBFailures"
      when e matches (botype == BUYCANC,originator == buyer,responder ==
store,outcome == tecFail)
            Cancellation in buyer.rights
      then
            if (Cancelation.BizFail == false)
                  then Cancelation.BizFail == true
                  else buyer reset
                        seller reset
            endif
end
```

Fig.18.

```
rule "Sample"
      when e matches (botype == BUYCANC,originator == buyer,responder ==
store)
            e.outcome == success
      then

end
```

Fig.19.

an outcome constraint. In the original version of EROP outcome constraints exist, but they are targeted at the event match conditions. For example, in the original specification of EROP the syntax in figure 19 was possible.

Where e is the event name and outcome is a property of the event. This made it possible to omit certain fields in the event match condition block and specify additional constraints after it. In the latest version of the implementation of Augmented Drools, an event match block requires all of the fields (type/originator/responder/status) to be specified. This defeats the purpose of the Outcome constraints as introduced in the original version of EROP and makes it so that it is no longer needed in the form that it was introduced; however, in the latest version of the ontology, as seen in the contract, outcome constraints can be used on business operations to check for example if the business operation has failed. This is expressed in clauses 2,4,6,8 and 10 of the contract presented in the previous section. The functionality to make such checks were not present in the

```
rule "Sample"
        when e matches (botype == BUYCANC)
        e.originator == buyer
        e.responder == store
        then

end
```

Fig.20.

original version of EROP, to accommodate it we have amended the original outcome constraint to have the following syntax and role.

$$BusinessOperation.BizzFail == true/false \qquad (1)$$

The construct can be used both in the Left-hand side and the right hand side of a rule, with the same syntax but a different meaning. When used in the left-hand side it is placed after the event match condition, the same way as it was introduced originally. The role when placed in the left-hand side of a rule is to check if the specified business action has happened, it other words it is a Boolean condition. When used in the right-hand side of a rule it servers not as a Boolean condition but rather a way to specify that the condition happened or did not happen. In other words when used in the right-hand side of a rule it serves as a setter. The change makes possible expressing clauses 2,4,6,8 and 10 and it updates the no longer needed version of outcome constraints as expressed in the original version of EROP in [5]. The requirement of the latest version of Augmented Drools that all the event match condition fields must be specified also makes the Role player constraint obsolete. The syntax in figure 20 is no longer needed and can be removed from the language.

The same as outcome constraints, role player constraints were used to add additional Boolean conditions after the event match block, given that some of the event match condition fields were omitted, given that the methods that were used to check for that functionality have been removed from the implementation of Augmented Drools it is no longer possible to do that.

### 5.2 Resetting Rop sets

The contract in Augmented Drools has another feature that is not present in the original specification of the EROP language. It is also captured by clauses 2,4,6,8 and 10 and it gives the ability to reset the ROP set of a given role player. This is needed to keep the consistency of the ROP sets of role players in the case of certain exceptional situations such as technical or business failures. To accommodate that functionality we have added the keyword reset to the grammar of EROP, which enables the contract writer to reset the ROP sets of a given role player. It can only be used in the right-hand side of a rule, similarly to ROP set manipulation. A sample of the operation is as follows in figure 21.

```
rule "CancellationBFailures"
    when e matches (botype == BUYCANC,originator == buyer,responder == store,outcome == tecFail)
    then
            seller reset

end
```

Fig.21.

### 5.3 Case study evaluation

When the contract, shown in the previous section is inputted in the Translator it produces the following results. A rule file in Augmented Drools, like one in EROP, starts with a declaration section where all the objects and entities used in the file are declared. After some java statements to import the classes of the EROP ontology, there is a section to declare global identifiers such as Role Players, Composite Obligations and Business Operations. Augmented Drools also needs instances of some other EROP ontology classes such as the Relevance Engine and the Event Logger for reference in the rules. The translator also automatically generates ROP sets for each Role Player specified in the declaration section (conforming by functional requirements FR1, FR2, FR3 and FR4). The translated declaration section looks as follows:

Package.  BuyerStoreContractEx
import    uk . ac . ncl . erop . ∗ ;

import    uk . ac . ncl . logging . CCCLogger ;

global    RelevanceEngine    engine ;    global    EventLogger    logger ;
global    RolePlayer    buyer ;

```
global    ROPSet       ropBuyer
global    RolePlayer   s e l l e r ;
global    ROPSet       ropSeller
global    BusinessOperation    buyRequest ;
global    BusinessOperation    payment ;
global    BusinessOperation    buyConfirm ;
global    BusinessOperation    buyReject ;
global    BusinessOperation    cancelation ;
```

The translator correctly generates instances of the Relevance Engine and Event Logger as well as the two Role Players and their corresponding ROP sets and all the specified Business Operations (Operations names start with lower case since they are java object and must follow Java style rules, as specified in FR5). The syntax to define rules in Augmented Drools is the same as in EROP given that the latter is derived from the former and has the following structure:

```
rule  RuleName  when  conditions  then
actions end
```

The Translator produces the following translation of the first two Rules (figure 22).

```
rule "BuyRequestReceived"
    when $e: Event(type=="BUYREQ", originator=="buyer", responder=="store",
status=="success")
        eval(ropBuyer.matchesRights(buyRequest))
    then
        ropBuyer.removeRight(buyRequest, seller);
        BusinessOperation[] bos = {buyConfirm, buyReject};
        ropSeller.addObligation("ReactToBuyRequest", bos, buyer,"01-01-
        2016 12:00:00"));
end
```

Fig.22.

The placeholder event variable is correctly translated to $e and the event match conditions are specified in the Augmented Drools format. Constraints on event

attributes is imposed outside the event match using the eval keyword as well as the methods from the Augmented Drools implementation (as specified in FR6). The right-hand side of the rule is translated correctly with manipulation of ROP sets going through method calls of the generated ROP sets of roleplayers. As expected, in the case of composite obligations, an extra line of code is needed to add a new composite obligation. In the above translation a composite obligation called bos is created and it consists of two other business operations. The second EROP rule, derived from clause 2 is translated to two rules in Augmented Drools because of the conditional structure used (figure 23).

The single rule in EROP is correctly broken down into two rules in Augmented Drools (as specified in FR7) the first one consisting of the conditions of the if statement added to the left-hand side of the rule and the then action added to the right hand sand of the rule. This rule also demonstrates the changed outcome constraints at work, correctly matching them as Boolean conditions and setters on the appropriate places (Lhs/Rhs). The second rule is produced by adding the negated conditions of the if statement to the left-hand side of the rule, while adding the then action to the right hand side of the rule. It also demonstrates the translation of the newly added reset construct that allows a contract writer to reset the ROP sets of a given role player.

The rest of the translation produces similar results, correct for the constructs used. The full translation is attached in the appendix section and can be compared against the original of the contract, which is also included, to verify its correctness.

## 6   Conclusions

The developed solution can generate translations, however there are improvements that can be made to enhance its capabilities. Some of these include: Adding more descriptive error handling mechanism, as currently, whenever the translator tries to parse a file, it expects the input to follow a certain format as specified in the grammar. If it does not find what it expects, error messages are presented that show the line of the file and character position at which the parser found an unexpected input. The error messages can be enhanced and made more descriptive and user friendly. Currently the translator is a stand-alone entity, separate from the CCC. It is worth exploring the cost of integrating it with the CCC and the amount of work required to do so. Another improvement would be enhancing the translator so that it can translate EROP to other Rules Languages. Even without any of the suggested features, the produced solution is capable with dealing with any of the EROP language constructs and their translations.

```
rule "BuyRequestBnessFailureIfThen"
    when $e: Event(type == "buyreq",originator == "buyer",responder ==
    "store",status == "tecfail")
        eval(buyRequest.getBusinessFailure() == false)
        eval(ropBuyer.matchesRights(BuyRequest))
    then
        buyRequest.setBusinessFailure == (true)
end

rule "BuyRequestBness1stFailureIfElse"
    when $e: Event(type == "buyreq",originator == "buyer",responder ==
    "store",status == "tecfail")
        eval(!buyRequest.getBusinessFailure() == false)
        eval(ropBuyer.matchesRights(BuyRequest))
    then
        ropBuyer.reset();
        ropSeller.reset();
end
```

Fig.23.